\documentclass[10pt]{article}

\usepackage{amsmath}
\usepackage{amssymb}
\usepackage{graphics}

\title{\textbf{Formalism Locality in Quantum Theory and Quantum Gravity}}

\author{Lucien Hardy\\
\textit{Perimeter Institute,}\\
\textit{31 Caroline Street North,}\\
\textit{Waterloo, Ontario N2L 2Y5, Canada}}

\begin{document}

\maketitle

\begin{abstract}
We expect a theory of Quantum Gravity to be both probabilistic and have indefinite causal structure.  Indefinite
causal structure poses particular problems for theory formulation since many of the core ideas used in the usual
approaches to theory construction depend on having definite causal structure.  For example, the notion of a state
across space evolving in time requires that we have some definite causal structure so we can define a state on a
space-like hypersurface.  We will see that many of these problems are mitigated if we are able to formulate the
theory in a {\it formalism local} (or F-local) fashion.  A formulation of a physical theory is said to be F-local
if, in making predictions for any given arbitrary space-time region, we need only refer to mathematical objects
pertaining to that region.  This is a desirable property both on the grounds of efficiency and since, if we have
indefinite causal structure, it is not clear how to select some other space-time region on which our calculations
may depend.  The usual ways of formulating physical theories (the time evolving state picture, the histories
approach, and the local equations approach) are not F-local.

We set up a framework for probabilistic theories with indefinite causal structure.  This, the causaloid framework, is
F-local.   We describe how Quantum Theory can be formulated in the causaloid framework (in an F-local fashion). This
provides yet another formulation of Quantum Theory.  This formulation, however, may be particularly relevant to the
problem of finding a theory of Quantum Gravity.
\end{abstract}

\section{Introduction}

The problem of Quantum Gravity is to find a theory which reduces in appropriate limits to General Relativity and
Quantum Theory (including, at least, those situations where those two theories have been experimentally confirmed).  To
be significant, the theory must also make correct predictions for new experiments in the future. The problem of
combining two less fundamental theories into a more fundamental one is not something for which a simple algorithm can
exist and thus we need a motivating idea to get started.  Here we note that General Relativity and Quantum Theory are
each conservative and radical in complementary ways. General Relativity is conservative in that it is deterministic but
radical in that it has non-fixed causal structure (whether a particular interval is time-like is not fixed in advance
but can only be decided after we have solved for the metric). Quantum Theory is conservative in that it has fixed
causal structure built in, but radical in that it is inherently probabilistic (standard Quantum Theory cannot be
formulated without reference to probabilities). It seems likely that a theory of Quantum Gravity must inherit the
radical features of the two component theories. Hence, we are looking for a theoretical structure that
\begin{enumerate}
\item  is probabilistic
\item  has non-fixed causal structure
\end{enumerate}
In fact, we expect the situation to be even more radical.  In General Relativity the causal structure is not fixed in
advance, but, once determined, there is a definite answer to the question of whether an interval is time-like or not.
However, in Quantum Theory any quantity that is subject to variation is also subject to quantum uncertainty.  This
means that, in a theory of Quantum Gravity, there may be no matter of fact as to whether a particular interval is
time-like or not. It is likely that the causal structure is not only non-fixed, but also indefinite.  The fact that we
expect the conservative features of each component theory to be replaced by the radical features in the other suggests
that a theory of Quantum Gravity cannot be entirely formulated within General Relativity or Quantum Theory. In this,
our program differs from String Theory \cite{string} and Loop Quantum Gravity \cite{loop} where the attempt is to
formulate Quantum Gravity within Quantum Theory (though there are other approaches which, to varying extents, do not
assume Quantum Theory will remain intact \cite{Penrose, Sorkin, Isham, Hartle}).

One signature of the fixed causal structure in Quantum Theory is the fact that we have a fixed background time $t$ used
to evolve the state $|\psi(t)\rangle = U(t)|\psi(0)\rangle$.  A deeper signature of fixed causal structure in Quantum
Theory can be seen by considering the different ways in which operators can be put together.   The operators
corresponding to two space-like separated regions are combined with the tensor product $A\otimes B$.  If a system
passes through two immediately sequential time-like separated regions then the appropriate way to combine the
corresponding operators is with the direct product $CB$. And if a system passes through two time-like separated regions
which have a gap in between (i.e. they are not immediately sequential) then the appropriate way to combine the
operators is with what we will call the question mark product $[D? B]$.  This linear operator is defined by
$[D?B]C\equiv DCB$. For each situation,  we must combine the associated operators in a way that depends on the causal
relationship between the two regions.  It would be good to have a mathematical framework which treats each type of
situation on an equal footing since then the fixed causal background need not be ingrained into the very structure of
the theory.

The task becomes one of finding a theoretical framework for probabilistic theories with indefinite causal structure
that correlate recorded data.  The causaloid framework set up in \cite{HardyQG1} (see also \cite{HardyQG2}) does this.
In this framework the {\it causaloid product} is defined.  This unifies the three products mentioned above from Quantum
Theory (in the context of a more general mathematical framework).

In this paper I will discuss the challenges posed by having indefinite causal structure and show how the causaloid
formalism deals with them.  I will indicate how the Quantum Theory of pairwise interacting qubits can be dealt with
in this formalism (this is an important example since we can use it to do universal quantum computation).  Finally
we will look at the road to formulating Quantum Gravity in this framework.

\section{Dealing with indefinite causal structure}

Indefinite causal structure is much more radical than merely having non-fixed causal structure as in General
Relativity (GR). In GR the causal structure, whilst not given in advance, is part of the solution.  After solving
Einstein's field equations we know the metric and therefore the causal structure.

Indefinite causal structure would mark a radical departure from previous physics. Many of our basic concepts and modes
of thought rely on having definite causal structure.  For example, we often think of quantities being conserved (in
time) or increasing (in time), and we think of information flowing (in time).  We think of entanglement (across space).
And, most crucially, we often think of a state (across space) evolving (in time). However, if we have indefinite causal
structure there would, in general, be no matter of fact as to whether a particular interval was space-like or time-like
and so all of these concepts and modes of thought would be placed under some tension. Nevertheless, one can make a very
strong case that Quantum Gravity (QG) will have indefinite causal structure and so we need to think sufficiently
radically to be able to be in a position to deal with this. Most approaches to QG do imagine some form of indefinite
causal structure. However, there has been comparatively little thought as to how to really deal with this properly.
Generally the conceptual and mathematical tools handed down to us from previous physics, encumbered with ingrained
notions of definite causal structure, are used.  For example, one might argue that we can model indefinite causal
structure by taking a sum over histories each having its own definite causal structure.  But why require each history
to have definite causal structure rather than giving up this notion all together at the fundamental level.  We need to
be prepared to think radically about this issue. The causaloid formalism offers a way forward here.

A common attitude is that the equations of physics must tell us how to calculate the evolution of physical systems {\it
in time}. If there is indefinite causal structure then we cannot think in this way.  Instead, we adopt the assertion
that {\it a physical theory must correlate recorded data}. This does not commit us to a picture of anything evolving in
time.  Thus, we might ask what
\begin{equation}\label{data12}
{\rm prob}({\rm data}_2|{\rm data}_1)
\end{equation}
is equal to.  If we can deal with all such probabilities for any data then we can say that we have formulated a
physical theory (at least that aspect of the theory which can be empirically verified).  By thinking about how data
might be correlated, we are adopting an operational methodology here. However, this is just a methodology aimed at
helping along theory construction.  In adopting this approach we do not commit ourselves to operationalism as a
fundamental philosophical outlook on the world.

We will now discuss two issues which arise when we think in this way (particularly when there is indefinite causal
structure).

\subsection{Issue 1: The need for a two-step approach}

The first issue is the question of when we have sufficient information to be able to make a prediction. Though this is
often not appreciated, physical theories only attempt to answer a very small fraction of the possible questions about
the world one might put to them. To see this, consider a spin-half particle subjected to three sequential spin
measurements. The probability that spin up is seen at the second position given that spin up was seen at the first
position, and given that the angles chosen were $\theta_1$ and $\theta_2$ (in the first and second positions) can be
written
\begin{equation}
{\rm prob}(+_2|+_1, \theta_1, \theta_2).
\end{equation}
This probability can be calculated using QT (and is equal to $\cos^2(\frac{\theta_2-\theta_1}{2})$).  We can say that
this probability is well defined.  This is an example of a question which the theory does answer.  But now consider the
probability that spin up is seen at the third position given that spin up was seen at the first position, and given
that the angles chosen were $\theta_1$ and $\theta_3$ in the first and third positions).  We can write this probability
as
\begin{equation}
{\rm prob}(+_3|+_1, \theta_1, \theta_3)
\end{equation}
Note that we are not given any information about the second spin measurement.  This is not part of the
conditioning. Under these circumstances we cannot use quantum theory to calculate this probability.  This
probability is not well defined.  This is a question which QT does not answer.   And neither should it.  Indeed,
generically QT does not answer most questions. This is true of physical theories in general (even deterministic
ones).  In General Relativity, for example, we can only make predictions about data that may be recorded in some
region $R_2$ given data in region $R_1$ if $R_2$ is a domain of dependence of $R_1$.

The key difference between the two situations in the spin example is to do with the causal structure.  In the first
case one measurement immediately proceeds the other, whereas in the second case there is a gap in time for which we
have no information.  In order to know whether the probability is well defined or not we need to know what causal
situation pertains.  If we have definite causal structure then we can refer to it and know whether we are in the
rather special type of situation where we can actually make a prediction.  However, if we have indefinite causal
structure then we do not know how to proceed.

No doubt there will still be certain conditional probabilities which are well defined even if go beyond quantum
theory and have indefinite causal structure. One way we might deal with this is to mathematize the question. Thus
we want a formalism involving two steps
\begin{quote}
{\bf Two step approach}
\begin{description}
\item[Step 1] We have a mathematical condition that is satisfied if and only if a probability is well defined.
\item[Step 2] In the case where the condition in step 1 is satisfied, we have a formula for calculating the probability.
\end{description}
\end{quote}
The standard picture with definite causal structure is actually an example of this form.  Thus, we have the theory of
domains of dependence which tell us whether we can make predictions about some region $R_2$ based on data in region
$R_1$ by looking at the causal structure. However, we can imagine more general ways in which we might implement this
two step approach that do not explicitly refer to causal structure (at least as the latter is usually conceived).

\subsection{Issue 2:  The need for F-locality}

The second issue is very much related to the first.  Imagine we want to calculate probabilities pertaining to some
arbitrary space-time region $R$.  This space-time region may be of any shape and may be disconnected (in so much as we
have a notion of connection in the absence of definite causal structure).  For example we may want to know what the
probability of seeing a certain outcome in $R$ is given that we performed certain measurements in $R$ and saw certain
other outcomes in $R$.  In the standard formulation of QT we have a state across space evolving in time. Imagine that
$R$ consists of two disconnected parts that are time-like separated. To make a prediction (to say whether the
probability is well defined and, if so, what it is equal to) we need to evolve the quantum state through intermediate
times.  Therefore we necessarily need to refer to mathematical objects and (implicitly) data which does not pertain to
$R$ in the evolving state picture.  As we will see, this is also the case in other types of formulation of physical
theories (such as histories approaches).  If we have some well defined causal structure then we can use that to tell us
what other region, besides $R$, we need to be considering to implement the mathematical machinery of the physical
theory. However, this option is not open to us if we have indefinite causal structure. In that case the only clean
approach is to insist that, in making predictions for $R$, we only refer to mathematical objects pertaining to $R$
(for, if not, what do we consider). This seems like a useful idea and deserves a name - we will call it formalism
locality (or F-locality).
\begin{quote}
{\bf F-locality}: A formulation of a physical theory is F-local if, in using it to make statements (using the two
step approach) about an arbitrary spacetime region $R$, we need only refer to mathematical objects pertaining to
$R$.
\end{quote}
It is possible that a given physical theory can be formulated in different ways.  F-locality is a property of the
way the physical theory is formulated rather than of the theory itself.  It is possible that any theory admits a
formulation which is F-local.  In the case where there is a definite causal structure we may be able to provide
both F-local and not F-local formulations of a theory.  However, if there is indefinite causal structure then it
seems likely that any fundamental formulation of the theory will necessarily be F-local.

There are two motivations for attempting to formulate theories in an F-local fashion:
\begin{enumerate}
\item We do not need to refer to some definite causal structure to decide what other region (besides the region
under consideration) to consider.
\item It is more efficient to consider only mathematical objects pertaining to the given region.
\end{enumerate}
Both these reasons are worth bearing in mind when evaluating formulations which are not F-local.

\section{How standard formulations of physical theories are not F-local}

There are, perhaps, three ways in which physical theories have been formulated to date.
\begin{enumerate}
\item The state evolving in time picture.
\item Histories formulations.
\item Local equations approach.
\end{enumerate}
None of these are F-local (the third case is a little more debatable) as we will now see.

In the state evolving in time picture the state is specified at some initial time and it then evolves according to
some equations.  Imagine we want to make a statement about a space-time region, $R$, consisting of two disconnected
parts that are separated in both space and time.  To do this we take a state defined across enough of space to
encompass both spatial regions and evolve it through enough time to encompass the two regions. Hence, we need to
refer to mathematical objects pertaining to a region of space-time $R'$ which includes both spatial and temporal
regions that are not part of $R$.

In histories formulations we consider the entire history from some initial to some final time. The physical theory
makes statements about such entire histories (the path integral formulation of Quantum Theory is one example).  If
we are only interested in some particular region $R$ then, clearly, in a history formulation, we need to make
reference to mathematical objects which do not pertain to $R$ and so the formulation is not F-local.  One might
claim that, since we can take the history across all of space-time, we do not need to refer to any definite causal
structure to decide what region to consider - we simply take everything.  Even if this does work, it is still more
efficient to aim at an F-local formulation. And, in practise, we always take our histories over some limited time
interval. Indeed, in the absence of a solution, we may not know the nature of \lq\lq all of space-time\rq\rq and so
it is difficult to know how to how to give a histories formulation of the theory.

An example of the local equations approach is Maxwell's equations.  Such equations constitute a set of local
statements about the infinitesimal regions making up our region $R$.  To actually make a prediction for region $R$
we need to use these local statements appropriately.  Typically this involves solving the equations with boundary
conditions on a boundary that is in the causal past of all of $R$. Hence, we need to consider a region bigger than
$R$. There may be other ways to utilize local equations to make predictions about arbitrary regions that do not
require consideration of a larger region.  It is clear, in any case, that a local equations formulation is not
explicitly F-local as defined above because it does not come equipped with an F-local technique for making
predictions for arbitrary regions. There is one sense in which local equations clearly go against the spirit of
F-locality.  A local equation relates quantities in regions that are infinitesimally displaced from one another.
The property of being infinitesimally displaced relates to causal structure. If the causal structure is indefinite
then it is not clear that we can retain this notion (this is one reason that we may expect whatever plays the role
of space time in a theory of Quantum Gravity to be discrete rather than continuous).

\section{An outline of the causaloid framework}

It is not clear that physical theories can be formulated in an F-local fashion.  In \cite{HardyQG1, HardyQG2} a
framework for probabilistic theories with indefinite causal structure was given.  This framework provides a way of
explicitly formulating theories in an F-local fashion.  Here we will give a bare-bones outline of this framework.
In the next section we will indicate how the QT of interacting qubits can be formulated in the framework.

\subsection{Data and regions}

Imagine that all the data collected during an experiment is recorded on cards as triples $(x, F_x, Y_x)$.  Here $x$
is some recorded data taken as representing space-time location, $F_x$ is some choice of experiment (knob setting)
at $x$, and $Y_x$ is the outcome of some experiment at $x$. For example, $x$ might be recorded from a GPS system,
$F_x$ could be the angle at which a Stern-Gerlach apparatus is set, and $Y_x$ could be the outcome of the spin
measurement.  During a typical experiment, data will be recorded at many space-time locations.  At the end of one
run of the experiment we will collect a stack of cards.  Since we are interested in probabilities we can imagine
running the experiment many times so we can obtain relative frequencies.

Since $x$ constitutes recorded data, it will be discrete.  Therefore, we can suppose that space-time is discrete
and comprised of elementary regions $R_x$.  We do not assume any a priori causal structure on the $x$.   An
arbitrary region $R_1$ consists of some set of elementary regions $R_x$
\begin{equation}
R_1 = \bigcup_{x\in{\mathcal O}_1} R_x
\end{equation}
We let $F_1$ denote the list of knob settings $F_x$ for $x\in{\mathcal O}_1$ and $Y_1$ denote the list of outcomes
for $x\in{\mathcal O}_1$.  $F_1$ denotes the choices made in $R_1$ and $Y_1$ denotes the outcomes in $R_1$
(sometimes we will use the longhand notation $F_{R_1}$ and $Y_{R_1}$ for $F_1$ and $Y_1$ respectively).

We can ask what
\begin{equation}
{\rm prob}(Y_2|Y_1, F_1, F_2)
\end{equation}
is equal to.  That is what is the probability of seeing outcomes $Y_2$ in region $R_2$ given that chose $F_1$ in
region $R_1$, chose $F_2$ in $R_2$, and saw outcomes $Y_1$ in region $R_1$ (this is a more sophisticated version of
(\ref{data12}))?  Here we are trying to make statements about region $R_1\cup R_2$.  We will now outline how we go
about doing this in an F-local way for $R_1\cup R_2$.

\subsection{${\bf p}$-type vectors and ${\bf r}$-type vectors}

Let $V$ be the union of all elementary regions.   We will assume that the probabilities
\begin{equation}
{\rm prob}(Y_V|F_V)
\end{equation}
are well defined (we are glossing over subtle points that are covered in \cite{HardyQG1}).  We can write
\begin{equation}
{\rm prob}(Y_V|F_V)= {\rm prob}(Y_{R_1}, Y_{V-R_1} |F_{R_1}, F_{V-R_1})
\end{equation}
We will now label each possible $( Y_{R_1}, F_{R_1})$ combination in region $R_1$ with $\alpha_1\in\Upsilon_1$.
This label runs over all possible $({\rm outcome}, {\rm choice})$ combinations in region $R_1$.  We write
\begin{equation}
p_{\alpha_1}= {\rm prob}(Y^{\alpha_1}_{R_1}, Y_{V-R_1} |F^{\alpha_1}_{R_1}, F_{V-R_1})
\end{equation}
We can regard $(Y_{V-R_1}, F_{V-R_1})$ in $V-R_1$ as a kind of generalized preparation for region $R_1$ (it is
generalized since it pertains to both the future and the past in so far as those concepts have meaning). Associated
with each generalized preparation is a state. We define the state for region $R_1$ to be that thing which is
represented by any mathematical object which can be used to calculate an arbitrary probability $p_{\alpha_1}$.
Clearly the object
\begin{equation}
\left( \begin{array}{c}  \vdots \\ p_{\alpha_1} \\ \vdots  \end{array} \right) ~~~~ \alpha_1\in\Upsilon_1
\end{equation}
suffices (since it simply lists all the probabilities).  However, in general, we expect that this is much more
information than necessary. In general, in physical theories, all quantities can be calculated from a subset of
quantities. We call this physical compression.  In our particular case we expect that a general probability
$p_{\alpha_1}$ can be calculated from a subset of these probabilities. We will restrict ourselves to linear
physical compression (where the probabilities are related by linear equations). We set
\begin{equation}
{\bf p} = \left( \begin{array}{c}  \vdots \\ p_{k_1} \\ \vdots  \end{array} \right) ~~~ k_1\in\Omega_1\subseteq
\Upsilon_1
\end{equation}
such that a general probability $p_{\alpha_1}$ can be calculated from the $p_{k_1}$ (with $k_1\in\Omega_1$) by a
linear equation
\begin{equation}
p_{\alpha_1} = {\bf r}_{\alpha_1} \cdot {\bf p}
\end{equation}
We chose the fiducial set $\Omega_1$ of labels such that there is no other choice with smaller $|\Omega_1|$ (this means
that every probability in ${\bf p}$ is necessary in the specification of the state).  There may be many possible
choices for the fiducial set $\Omega_1$.  We simply choose one (for each region) and stick with it.  We do not lose
generality by imposing linearity here. In the worst case $\Omega_1=\Upsilon_1$. It is possible that nonlinear
compression is more efficient. However, for probabilities this is not the case as long as one can form arbitrary
mixtures. In particular, in Quantum theory (and Classical Probability Theory) linear compression is optimal (so long as
we allow mixed states rather than restrict ourselves to pure states).

Associated with each region $R_1$ is a real vector space of dimension $|\Omega_1|$.  Further, associated with each
$(Y_1, F_1)$ combination there is a ${\bf r}$-type vector (which lives in a dual space to the ${\bf p}$-type
vectors representing the states) which we can write as
\begin{equation}
{\bf r}_{(Y_1, F_1)}(R_1)
\end{equation}
or ${\bf r}_{\alpha_1}$ for short.  It is these ${\bf r}$-type vectors that we use in real calculations.  The state
vector, ${\bf p}$, is akin to scaffolding - it can be dispensed with once the structure of the ${\bf r}$-type vectors
is in place as we will see shortly.

\subsection{The causaloid product}

If we have two disjoint regions $R_1$ and $R_2$, then we can consider the region $R_{12}\equiv R_1\cup R_2$ as a
region in its own right. We can denote the outcome and knob settings for $R_{12}$ as $Y_1\cup Y_2$ and $F_1\cup
F_2$ (perhaps we are slightly abusing the $\cup$ notation here).  Since $R_1\cup R_2$ is a region in its own right
we will have ${\bf r}$-type vectors associated with each $({\rm outcome}, {\rm choice})$ combination in this region
also.

We can label the $({\rm outcome}, {\rm choice})$ pairs $(Y_1\cup Y_2, F_1\cup F_2)$ with $\alpha_1\alpha_2\in
\Upsilon_1\times\Upsilon_2$ (where $\times$ denotes the cartesian product of ordered pairs taken from the two
sets). We also have the fiducial set $\Omega_{12}$ for this region.  There is an important theorem - namely that it
is always possible to choose $\Omega_{12}\subseteq \Omega_1\times\Omega_2$. We will use the notation $l_1l_2\in
\Omega_1\times\Omega_2$ and $k_1k_2\in \Omega_{12}$.

In the case that $\Omega_{12} = \Omega_1\times\Omega_2$ we have no extra physical compression when two regions are
considered together.  However, if $\Omega_{12}\subset \Omega_1\times\Omega_2$ then there is an extra physical
compression (second level compression) for the composite region over and above the physical compression (first level
compression) for each component region considered separately.  Physically, this non-trivial case corresponds to {\it
causal adjacency} such as when a qubit passes through two sequential regions with no gap in between.

Second level compression is quantified by a matrix $\Lambda_{l_1l_2}^{k_1k_2}$ (which depends on the composite
region under consideration) such that
\begin{equation}\label{defcausaloidprod}
{\bf r}_{\alpha_1\alpha_2}\big|_{k_1k_2} = \sum_{l_1l_2\in\Omega_1\times\Omega_2}  {\bf r}_{\alpha_1}\big|_{l_1}
{\bf r}_{\alpha_2}\big|_{l_2} \Lambda_{l_1l_2}^{k_1k_2}
\end{equation}
where ${\bf r}_{\alpha_1}\big|_{l_1}$ denotes the $l_1$ component of ${\bf r}_{\alpha_1}$.  We write
\begin{equation}
{\bf r}_{\alpha_1\alpha_2} = {\bf r}_{\alpha_1} \otimes^{\Lambda} {\bf r}_{\alpha_2}
\end{equation}
where the components are given in (\ref{defcausaloidprod}). Accordingly, we have defined a new type of product denoted
by $\otimes^{\Lambda}$.  We call this the {\it causaloid product}.  It is the sought after unification of the various
products $A\otimes B$, $AB$, and $[A?B]$ from quantum theory mentioned in the introduction (though in a more general
framework).  The form of the product is the same regardless of the causal structure.  However, since the $\Lambda$
matrix can differ for different composite regions, we can still encode the different products in this one product.

If we have a region regarded as being composed of more than two regions then we can generalize the above ideas
appropriately (calling on, in general, a matrix of the form $\Lambda_{l_1l_2l_3 \dots}^{k_1k_2k_3\dots}$).

\subsection{The two-step approach in the causaloid framework}

We note that, using Bayes rule,
\begin{eqnarray}
{\rm Prob}(Y_2|Y_1, F_1, F_2)&=& \frac{{\rm Prob}(Y_2,Y_1| F_1, F_2)}{\sum_{X_2\sim F_2} {\rm Prob}(X_2,Y_1| F_1,
F_2)}
\\
& &
\\
&=& \frac{{\bf r}_{(Y_2\cup Y_1, F_1\cup F_2)}\cdot {\bf p}}{\sum_{X_2\sim F_2} {\bf r}_{(X_2\cup Y_1, F_1\cup
F_2)}\cdot {\bf p}}
\end{eqnarray}
where the notation $X_2\sim F_2$ denotes all outcomes in $R_2$ which are consistent with the choice $F_2$ in $R_2$. In
order that the probability ${\rm Prob}(Y_2|Y_1, F_1, F_2)$ be well defined, it must depend only on the given
conditioning in $R_1\cup R_2$ and not depend on what happens outside this region.  The state, on the other hand, is
associated with some generalized preparation in $V-R_1-R_2$. Hence, this probability is only well defined if there is
no dependence on the state ${\bf p}$ (strictly we should have included the conditioning in $V-R_1-R_2$ and then shown
that it is irrelevant if there is no dependence on ${\bf p}$).  We can use this observation to implement a two step
approach.
\begin{quote}
{\bf Two step approach}
\begin{enumerate}
\item ${\rm Prob}(Y_2|Y_1, F_1, F_2)$ is well defined if and only if
\begin{equation}
{\bf r}_{(Y_2\cup Y_1, F_1\cup F_2)} \text{ is parallel to} \sum_{X_2\sim F_2} {\bf r}_{(X_2\cup Y_1, F_1\cup F_2)}
\end{equation}
(since this is the necessary and sufficient condition for there being no dependence on ${\bf p}$).
\item If these vectors are parallel, then the probability is given by
\begin{equation}
{\rm Prob}(Y_2|Y_1, F_1, F_2) = \frac{|{\bf r}_{(Y_2\cup Y_1, F_1\cup F_2)}|}{|\sum_{X_2\sim F_2} {\bf r}_{(X_2\cup
Y_1, F_1\cup F_2)}|}
\end{equation}
\end{enumerate}
\end{quote}
We see that this two-step approach does not require us to refer to some given definite causal structure (at least
as the latter is usually conceived).

We see that the formulation is F-local since, to make predictions for an arbitrary space-time region (in this case
the region $R_1\cup R_2$), we need only consider mathematical objects pertaining to this region (the ${\bf r}$
vectors). Strictly speaking, we need to be sure we can calculate the ${\bf r}$ vectors for arbitrary regions
without referring to mathematical objects pertaining to other regions to assert that the formulation is fully
F-local.  This will be addressed below.

\subsection{The causaloid}

We can calculate ${\bf r}$ vectors for an arbitrary region by starting from the ${\bf r}$ vectors for the
elementary regions comprising that region and using the causaloid product.   Hence, we can calculate any ${\bf
r}$-vector if we know
\begin{enumerate}
\item All the vectors ${\bf r}_{\alpha_x}$ (which can be regarded as a matrix $\Lambda_{\alpha_x}^{k_x}$) for all
elementary regions, $R_x$.
\item All the matrices $\Lambda_{l_xl_{x'}l_{x''} \dots}^{k_xk_{x'}k_{x''}\dots}$  with $x, x', x'', \dots
\in{\cal O}_1$,  for all ${\cal O}_1$ with $|{\cal O}_1|\geq 2$ (since these pertain to composite regions).
\end{enumerate}
This constitutes a tremendous amount of information (the number of matrices is exponential in the number of
elementary regions and the size of these matrices grows with the size of the region they pertain to). However, we
can apply physical compression by finding relationships between these matrices (we call this third level
compression). After applying physical compression, we have a smaller set of matrices from which all the others can
be calculated. We call this smaller set, augmented by a set of rules for implementing decompression, {\it the
causaloid} and denote it by $\bf \Lambda$.  Since we want the framework to be F-local, we require that, in applying
decompression to obtain the matrix for a region $R_1$, we only need use matrices pertaining to regions
$\tilde{R}_1\subseteq R_1$.

While we have not shown how to calculate the causaloid in general, it has been shown how to do so for the classical
probabilistic theory of pairwise interacting classical bits and the quantum theory of pairwise interacting qubits.
We will outline, in the next section, how this works in the quantum case.  The classical case is very similar
(though we will not outline it in this paper).

\section{Formulating Quantum Theory in the causaloid framework}

In this section we will content ourselves with simply describing how the Quantum Theory of pairwise interacting
qubits can be formulated in the causaloid framework without deriving any of the results.  Universal quantum
computation can be implemented with pairwise interacting qubits and so arbitrary quantum systems can be simulated
to arbitrary accuracy (similar comments apply in the classical case).  Hence the case we are studying is more than
just an example.  It demonstrates (with some appropriate qualifications) that Quantum Theory in general can be
formulated in this framework.

Assume we have a large number of qubits moving to the right labelled (from left to right) $u=1, 2, 3, \dots$ and a
large number of moving to the left labelled (from right to left) $v=1, 2, 3, \dots$.  If this is plotted against
time then we will have a diamond shaped lattice with each vertex corresponding to the interaction of a right moving
qubit with a left moving qubit.  We can label these vertices with $x\equiv uv$ (the cartesian product of $u$ and
$v$ is denoted by $uv$). They correspond to our elementary regions $R_{uv}$.

We imagine that, at each vertex, the two qubits pass through a box which implements a general measurement. The box
has a knob which is used to set $F_{uv}$ and a display panel recording the outcome, $Y_{uv}$. As before, we can
label all such pairs with $\alpha_{uv}$. For a general quantum measurement, an $(\text{outcome}, \text{choice})$
pair is associated with a superoperator $\$ $ (superoperators are trace non-increasing maps on density operators
that take allowed states to allowed states).  In this case, we have a superoperator $\$_{\alpha_{uv}}$ associated
with $\alpha_{uv}$. In quantum theory we can write a general superoperator on two qubits such as this as a sum of
the tensor product of a fiducial set of superoperators acting on each qubit separately:
\begin{equation}\label{qelambda}
\$_{\alpha_{uv}} = \sum_{k_uk_v\in\Omega^2\times\Omega^2} \Lambda_{\alpha_{uv}}^{k_uk_v} \$_{k_u}\otimes \$_{k_v}
\end{equation}
Remarkably, we can only do this if we have complex (rather than real or quaternionic) Hilbert spaces supporting the
superoperators. The set $\$_{k_u}$ for $k_u\in\Omega^2$ is a fiducial spanning set of superoperators for the qubit (the
superscript denotes that this is a qubit having Hilbert space dimension 2).  We have $|\Omega^N|=N^4$ so for a qubit we
have $|\Omega|^2=2^4$. This is the number of linearly independent superoperators needed to span the space of
superoperators for a qubit.  For reasons that will be clear later, we choose $ \$_1 = I$ where $1$ is the first element
of $\Omega^2$ and $I$ is the identity superoperator. We can solve equation (\ref{qelambda}) to find the $\Lambda$
matrix for each elementary region $R_{uv}$.

Now consider a single right moving qubit as it goes from vertex $(u,v)$ to the next vertex $(u,v+1)$.  Assume that
it is subject to $ \$_{l_u}\otimes \$_{l_v}$ at the first vertex and $\$_{l'_u}\otimes \$_{l'_{v+1}}$ at the next
with $l_u, l'_u, l_v, l'_{v+1} \in \Omega^2$.  So far as the right moving qubit, $u$, is concerned, we can ignore
the left moving qubits with which it interacts (since the two superoperators just given factorize).  The effective
superoperator acting on qubit $u$ is $\$_{l'_u}\circ\$_{l_u}$.  But this is a superoperator belonging to the space
of superoperators acting on a single qubit and can, hence, be expanded in terms of the linearly independent
fiducial set
\begin{equation}\label{qplambda}
\$_{l'_u}\circ\$_{l_u} = \sum_{k'_uk_u\in\{1\}\times\Omega^2} \Lambda_{l'_ul_u}^{k'_uk_u} \$_{k'_u}\circ\$_{k_u}
\end{equation}
since we have selected $\$_1$ to be the identity.  We can solve this this equation for the matrix
$\Lambda_{l'_ul_u}^{k'_uk_u}$.

This generalizes to more than two sequential vertices in the obvious way. For three sequential vertices we have
\begin{equation}
\$_{l''_u}\circ\$_{l'_u}\circ\$_{l_u}   = \sum_{k''_uk'_uk_u\in\{1\}\times\{1\}\times\Omega^2}
\Lambda_{l''_ul'_ul_u}^{k''_uk'_uk_u} \$_{k''_u}\$_{k'_u}\circ\$_{k_u}
\end{equation}
and so on.

It can be shown that, for three sequential vertices,
\begin{equation}\label{compident}
\Lambda_{l''_ul'_ul_u}^{k''_uk'_uk_u} = \sum_{n'\in\Omega^2} \Lambda_{l''_ul'_u}^{k''_un'_u}\Lambda_{n'_ul_u}^{k'_uk_u}
\end{equation}
For four sequential vertices,
\begin{equation}\label{fourv}
\Lambda_{l'''_ul''_ul'_ul_u}^{k''_uk'_uk_u} = \sum_{n''\in\Omega^2 ~ n'\in\Omega^2} \Lambda_{l'''_ul''_u}^{k'''_un''_u}
\Lambda_{n''_ul'_u}^{k''_un'_u}\Lambda_{n'_ul_u}^{k'_uk_u}
\end{equation}
and so on.  The derivation of these equations relies only on the combinatorics of how the labels combine rather than on
any particular details of quantum theory.  The same equations are found in the treatment of interacting classical bits.

We may have need of the matrix $\Lambda_{l_u}^{k_u}$ (where $l_u, k_u\in\Omega^2$) for a single vertex for a right
moving qubit. Since
\begin{equation}
\$_{l_u} = \sum_{k_u\in\Omega^2} \Lambda_{l_u}^{k_u} \$_{k_u}
\end{equation}
we have
\begin{equation}\label{onev}
\Lambda_{l_u}^{k_u}=\delta_{l_u}^{k_u}
\end{equation}
For left moving qubits we have equations (\ref{qplambda}-\ref{onev}) but with $v$ replacing $u$.

As will be described, the composite region $\Lambda$ matrices for all situations that do not involve sequential
vertices on the same qubit are given by multiplying the $\Lambda$ matrix components for different clumps of
vertices. This means that the causaloid is given by
\begin{equation}\label{qcausaloid}
{\bf \Lambda} = ( \Lambda_{\alpha_{uv}}^{k_uk_v} \forall uv, \Lambda_{l'_ul_u}^{k'_uk_u} \forall \text{ RSV},
\Lambda_{l'_vl_v}^{k'_vk_v} \forall \text{ LSV}; \text{clumping method} )
\end{equation}
where (L)RSV stands for pairs of sequential vertices on (left) right moving qubits.  The clumping method allows us
to calculate the $\Lambda$ matrix for an arbitrary region $R_1$ with $x (=uv) \in {\cal O}_1$ from this causaloid
as follows
\begin{enumerate}
\item  For each qubit (both left and right moving) circle all complete groups of sequential vertices (call these
{\it clumps}) in ${\cal O}_1$. There must be a gap of at least one vertex between each clump for any given qubit.
\item Calculate the $\Lambda$ matrix components for each circled clump for each qubit using
(\ref{compident}) or one of its generalizations (for a clump of one vertex use (\ref{onev}) and for a clump of two
vertices take $\Lambda_{l'_ul_u}^{k'_uk_u}$ directly from the specification of the causaloid (\ref{qcausaloid})).
\item Multiply together all $\Lambda$ matrix components for all circled clumps (note we are
multiplying components rather than performing matrix multiplication).  This gives the components of the $\Lambda$
matrix for $R_1$.
\end{enumerate}
We note that the clumping method respects F-locality since, in calculating the $\Lambda$ matrix for $R_1$, we only
use $\Lambda$ matrices pertaining to regions $\tilde R_1 \subseteq R_1$.

It is worth examining the causaloid given in (\ref{qcausaloid}) a little more.  First we note that we are only
required to specify a tiny subset of the exponential number of possible $\Lambda$ matrices - there is a tremendous
amount of third level compression. Second, we note the symmetry property that, according to (\ref{qelambda}) and
(\ref{qplambda}), the $\Lambda$ matrices of each type are the same. Hence, we can actually specify the causaloid by
\begin{equation}
{\bf \Lambda} = ( \Lambda_{\alpha_{11}}^{k_1k_1}, \Lambda_{l'_1l_1}^{k'_1k_1}; \text{symmetry, clumping method} )
\end{equation}
where where {\it symmetry}  denotes the property just noted and $\Lambda_{l'_1l_1}^{k'_1k_1}$ is one instance of
the $\Lambda$ matrix for a pair of right (or left) sequential vertices.

Given this causaloid we can employ the standard techniques of the causaloid framework (the causaloid product and
the two step approach) to calculate whether an arbitrary probability is well defined and, if so, what it is equal
to.  In this sense we can say that this causaloid fully specifies the quantum theory of interacting qubits.  In
particular, note that we separate out the specification of the theory (the causaloid will be different for
different physical theories) from the way the causaloid is used to make predictions (it is used in the same way for
any physical theory).

The causaloid formulation of Quantum Theory treats arbitrary regions on an equal footing.  In this it is similar to the
time-symmetric approach of Aharanov and co-workers (particularly the latest version due to Aharanov, Popescu,
Tollaksen, and Vaidman in \cite{Aharanov} which allows states and measurements to be defined for arbitary regions), and
the general boundary formulation of quantum theory due to Oeckl \cite{Oeckl}.  Of related interest is the quantum
causal histories approach of Markopoulou \cite{Fotini} and the quantum causal networks of Leifer \cite{Leifer}.

\section{The road to Quantum Gravity}

In order to formulate a theory of Quantum Gravity (QG) we need to have a framework that is hospitable to such a
theory in the first place.  We expect that QG will be a probabilistic theory with indefinite causal structure. The
causaloid framework admits such theories (this does not imply that QG certainly fits in the framework - but at
least it is not ruled out from the outset).  The most satisfying way to obtain QG in this (or any) framework would
be to derive it from a set of well motivated principles (such as a suitably generalized equivalence principle).  It
may be that appropriate principles will carry over from Quantum Theory and General Relativity (indeed the
equivalence principle is true in Newtonian Gravity).  Therefore, a careful study of Quantum Theory  and General
Relativity (GR) may be the best way of coming up with such principles.

One particular route that may be taken to finding QG is illustrated in the following diagram
\[
\begin{array}{ccc}
  {\rm QT} & \longrightarrow & {\rm QG} \\
  \uparrow &  & \uparrow \\
  {\rm CProbT} & \longrightarrow & {\rm ProbGR}
\end{array}
\]
CProbT is classical probability theory (of interacting classical bits for example). QT is Quantum Theory (of
interacting qubits for example).  ProbGR is an appropriately formulated version of General Relativity in the case
where we have arbitary probabilistic ignorance of the values of certain measurable quantities.  We will elaborate
on this below.  The vertical arrows represent a kind of {\it quantization}. The horizontal arrows represent what we
might call {\it GR-ization}. In quantizing from CProbT to QT we need only alter the structure of the
$\Lambda_{\alpha_{uv}}^{k_uk_v}$ matrices for the elementary regions (this is what might be regarded as the local
structure) by replacing that structure that corresponds to a classical probability simplex with a structure that
corresponds to the Bloch sphere of the qubit. The structure above this, for composite regions, is basically
constructed in the same way in CProbT and QT.  ProbGR has not yet been satisfactorily formulated.  However, we can
expect that in the GR-ization process from CProbT to ProbGR, the local structure associated with the classical
probability simplices for will survive for the elementary regions but that the structure associated with composite
regions will be different (since the causal structure is not fixed).  This suggests that we may be able to get a
theory of QG by applying quantization essentially at the local level of the elementary regions and GR-ization at
the level of the composite regions.  If quantization and GR-ization do not interfere with each other too much then
the diagram above may not be too misleading.

This approach depends on having a suitable formulation of ProbGR.  An obvious way to give a probabilistic
formulation of GR is to have some probabilistic distribution over the 3-metric specified on some initial space-like
hypersurface and then evolve the distribution employing a canonical formulation of GR.  This is unsatisfactory
since (a) it is not F-local, (b) we cannot deal with arbitrary probabilistic ignorance about measurable quantities,
and (c) the time label for the space-like hypersurface is not an observable and so it is not clear that the numbers
we are calling \lq\lq probabilities\rq\rq represent ignorance about something measurable.  Another approach is to
consider a probabilistic distribution over solutions for the metric for all of space-time.  This is problematic
since (a) it is manifestly not F-local and (b) it is not clear how the differently weighted solutions match up from
the internal point of view of somebody who may be making measurements and so, once again, it is unclear whether the
\lq\lq probabilities\rq\rq represent ignorance about something measurable.  Rather than taking either of these
approaches, it seems that we need to build up ProbGR from scratch using F-locality and, maybe, the causaloid
formalism as guidance.  Such a theory contains no new empirical content over standard GR.  However, it is possible
that the natural mathematical formulation of ProbGR will look very different from standard GR.

\section{Conclusions}

Many standard notions in Quantum Theory require reference to some definite causal structure.  For example the
notion of entanglement requires two space-like separated systems, and the notion of information flow requires a
sequence of immediately sequential time-like regions.  When we embed QT into the causaloid framework these notions
become special cases of a much richer structure. Entanglement is supported by the tensor product of QT, but in the
causaloid framework, we have the causaloid product which allows us to talk about joint properties of any two
regions regardless of their causal relationship.  Information flow is supported by the standard product $\hat{
A}\hat{ B}$ between sequential time-like separated regions.  In the causaloid framework we have, again, the
causaloid product.  In quantum circuit diagrams we draw wires between boxes denoting the path of the qubit. A pair
of boxes either do, or do not, have a wire between them.  In the causaloid framework we have a $\Lambda$ matrix (by
which the causaloid product is defined). Every pair of boxes (or elementary regions) has a $\Lambda$ matrix between
it. This richer structure is likely to help in developing a theory of Quantum Gravity since it provides a way round
requiring that the causal structure be definite.

The principle that it should be possible to give an F-local formulation may prove to be powerful in theory
construction.  It is encouraging that Quantum Theory can be formulated in an F-local fashion.  Not only does this add
to the list of different ways in which QT can be formulated but also it provides encouragement that a theory of QG may
share structural similarities with QT.

The next step on the road to QG, if this approach is pursued, is to construct ProbGR.  In tackling the problem of
constructing ProbGR we are likely to encounter many of the same difficulties encountered in constructing a fully
fledged theory of QG. However, we know that ProbGR is empirically equivalent to GR (just with arbitrary probabilistic
ignorance added) and so we fully expect that this theory exists.

\end{document}